\documentclass[reprint,aip,twocolumn,cha,showkeys,showpacs,superscriptaddress]{revtex4-1}

\usepackage{amsmath,amssymb,color,bm}
\usepackage{graphicx}
\usepackage{ulem}
\renewcommand{\thetable}{\Roman{table}}
\renewcommand{\thesection}{\Roman{section}}
\usepackage[caption=false]{subfig}
\usepackage{booktabs}

\begin{document}
\title{Circuit Electromechanics with a Non-Metallized Nanobeam}

\author{M.~Pernpeintner}
\email{matthias.pernpeintner@wmi.badw.de}
\affiliation{Walther-Mei{\ss}ner-Institut, Bayerische Akademie der Wissenschaften, D-85748 Garching, Germany}
\affiliation{Nanosystems Initiative Munich, Schellingstra{\ss}e 4, D-80799 M\"{u}nchen, Germany}
\affiliation{Physik-Department, Technische Universit\"{a}t M\"{u}nchen, D-85748 Garching, Germany}
\author{T.~Faust}
\affiliation{Center for NanoScience (CeNS) and Fakult\"{a}t f\"{u}r Physik, Ludwig-Maximilians-Universit\"{a}t, D-80799 M\"{u}nchen, Germany}
\author{F.~Hocke}
\affiliation{Walther-Mei{\ss}ner-Institut, Bayerische Akademie der Wissenschaften, D-85748 Garching, Germany}
\affiliation{Nanosystems Initiative Munich, Schellingstra{\ss}e 4, D-80799 M\"{u}nchen, Germany}
\affiliation{Physik-Department, Technische Universit\"{a}t M\"{u}nchen, D-85748 Garching, Germany}
\author{J.\,P.~Kotthaus}
\affiliation{Center for NanoScience (CeNS) and Fakult\"{a}t f\"{u}r Physik, Ludwig-Maximilians-Universit\"{a}t, D-80799 M\"{u}nchen, Germany}
\author{E.\,M.~Weig}
\affiliation{Center for NanoScience (CeNS) and Fakult\"{a}t f\"{u}r Physik, Ludwig-Maximilians-Universit\"{a}t, D-80799 M\"{u}nchen, Germany}
\affiliation{Department of Physics, University of Konstanz, D-78457 Konstanz, Germany}
\author{H.~Huebl}
\email{hans.huebl@wmi.badw.de}
\affiliation{Walther-Mei{\ss}ner-Institut, Bayerische Akademie der Wissenschaften, D-85748 Garching, Germany}
\affiliation{Nanosystems Initiative Munich, Schellingstra{\ss}e 4, D-80799 M\"{u}nchen, Germany}
\author{R.~Gross}
\affiliation{Walther-Mei{\ss}ner-Institut, Bayerische Akademie der Wissenschaften, D-85748 Garching, Germany}
\affiliation{Nanosystems Initiative Munich, Schellingstra{\ss}e 4, D-80799 M\"{u}nchen, Germany}
\affiliation{Physik-Department, Technische Universit\"{a}t M\"{u}nchen, D-85748 Garching, Germany}

\date{\today}

\begin{abstract}
We have realized a nano-electromechanical hybrid system consisting of a silicon nitride beam dielectrically coupled to a superconducting microwave resonator. We characterize the sample by making use of the Duffing nonlinearity of the strongly driven beam. In particular, we calibrate the amplitude spectrum of the mechanical motion and determine the electromechanical vacuum coupling. A high quality factor of $480{,}000$ at a resonance frequency of $14\,\mathrm{MHz}$ is achieved at 0.5\,K. The experimentally determined electromechanical vacuum coupling of $11.5\,\mathrm{mHz}$ is quantitatively compared with finite element based model calculations.
\end{abstract}

\maketitle
In the field of cavity optomechanics, micro- or nanoscale mechanical resonators are coupled to an optical cavity allowing to transfer information from the mechanical to the optical domain and vice versa \cite{kippenberg_analysis_2005,gigan_self-cooling_2006,arcizet_radiation-pressure_2006, kleckner_sub-kelvin_2006,thompson_strong_2008,aspelmeyer_cavity_2014}.
For example, such optomechanical systems have recently been used to observe interference effects like optomechanically induced transparency \cite{weis_optomechanically_2010} or to cool mechanical resonators to their ground state \cite{chan_laser_2011}. Moreover, they have been e.\,g.~proposed for quantum information processing \cite{stannigel_optomechanical_2012} or for the detection of gravitational waves \cite{kaltenbaek_macroscopic_2012}. To increase the performance of such systems, high mechanical quality factors and high resonance frequencies are desirable, which are provided e.\,g.~by strongly pre-stressed Si$_3$N$_4$ thin films \cite{verbridge_high_2006,unterreithmeier_universal_2009,yu_control_2012}. Analogous to optical cavities, electrical circuits can be coupled to nanomechanical resonators giving rise to the field of cavity nano-electromechanics \cite{regal_measuring_2008,oconnell_quantum_2010,rocheleau_preparation_2010,teufel_circuit_2011,teufel_sideband_2011,zhou_slowing_2013}. Typically, a metallized mechanical resonator is capacitively coupled to a superconducting microwave resonator enabling e.\,g. ground state cooling \cite{teufel_circuit_2011} or the control of microwave signals \cite{zhou_slowing_2013}. In order to enable a further increase of quality factors it is beneficial to avoid any additional dissipation due to the metallization of the mechanical resonator \cite{beil_observation_2006,collin_metallic_2010,das_design_2012}. To this end, the dielectric coupling of non-metallized Si$_3$N$_4${} nanoresonators has been established as an alternative transduction and control scheme for high-Q nano-electromechanical systems in recent years \cite{unterreithmeier_universal_2009,rieger_frequency_2012}. Here, the displacement is detected dielectrically via a microwave cavity's resonance frequency which immediately enables cavity electromechanics \cite{faust_microwave_2012}. For the temperature range between 10 and 300\,K, conventional copper microstrip cavities have been successfully employed. Millikelvin temperature operation can be achieved by integrating a pure, i.\,e.~non-metallized, nanomechanical beams into a superconducting microwave circuit. This approach is promising for the realization of high quality factors of both the mechanical and the microwave resonator.

In this letter we demonstrate such an integrated nano-electromechanical system where a pure insulating silicon nitride nanobeam is coupled to a superconducting coplanar waveguide microwave resonator. We calibrate the mechanical amplitude of the beam and use its nonlinear Duffing response to obtain the electromechanical coupling (cf.~ref.~\citenum{hocke_exploring_2014}). We find a vacuum coupling of $g_0/2\pi=11.5\,\mathrm{mHz}$ which is corroborated quantitatively by finite element modelling of the system.

\begin{figure}[tb]
\centering
\includegraphics[width=0.99\columnwidth]{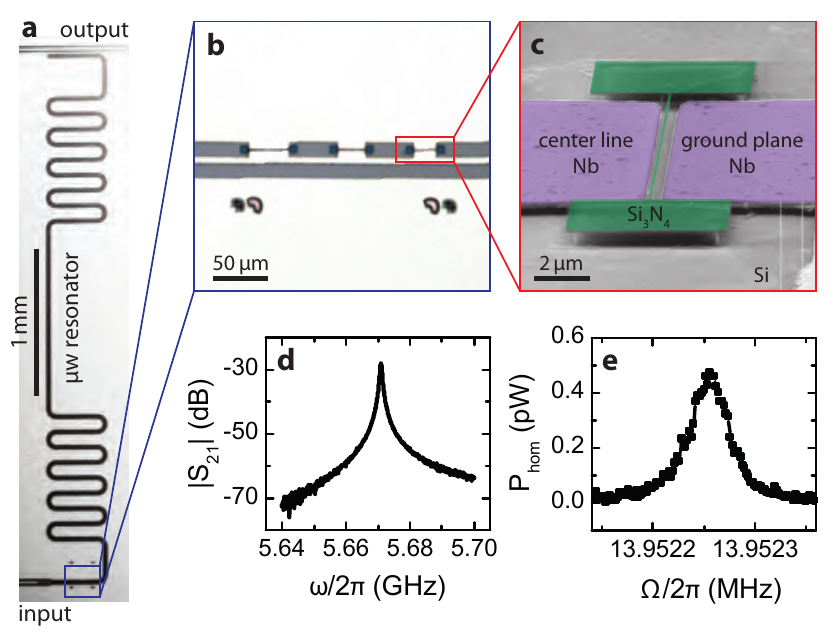}
\caption{\textbf{a-c.} Optical micrograph and false color scanning electron micrograph of the nano-electromechanical hybrid sample. \textbf{d.} Uncalibrated transmission spectrum of the superconducting microwave resonator. \textbf{e.} Driven power spectrum of the nanomechanical beam, measured with the homodyne setup depicted in Fig.~\ref{fig:CryoSetup-Homodyne-DWP-CalTone}.}
\label{fig:SEMimage-MicrowaveResonator}
\end{figure}

\begin{figure}[tb]
\centering
\includegraphics[width=0.95\columnwidth]{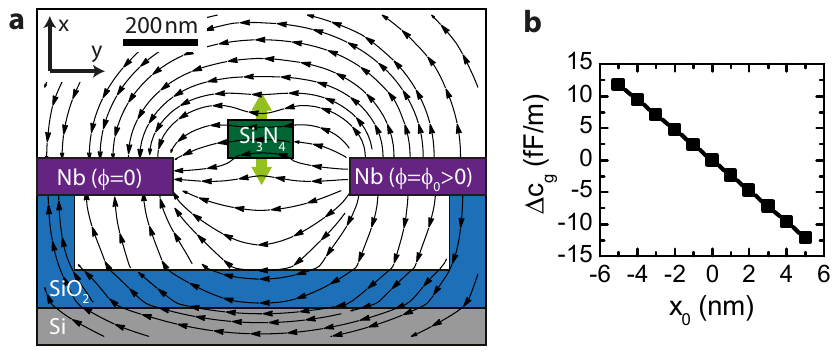}
\caption{\textbf{a.} Two-dimensional static COMSOL model to simulate the electromechanical coupling. The black arrows indicate the electric field strength and direction, resulting from an electrostatic potential difference $\phi_0$ between the niobium electrodes. The light green arrow illustrates the dynamic displacement of the nanobeam (not to scale). \textbf{b.} Calculated capacitance change per length, $\Delta c_g(x_0) \equiv c_g(x_0)-c_g(0)$, as a function of the beam displacement $x_0$.}
\label{fig:grAssembly-CapacityVsBeamDisplacement}
\end{figure}

Our nano-electromechanical device consists of a doubly-clamped, highly tensile-stressed Si$_3$N$_4${} nanobeam which is dielectrically coupled to a $\lambda/2$ superconducting niobium microwave resonator. The beam is located between the niobium centerline and ground plane with a gap of $150\,\mathrm{nm}$ between beam and electrodes, as shown in Fig.~\ref{fig:SEMimage-MicrowaveResonator}a-c. The sample is fabricated on a single-crystalline silicon wafer coated with $400\,\mathrm{nm}$ of thermal oxide and $100\,\mathrm{nm}$ of highly tensile-stressed LPCVD silicon nitride (Si$_3$N$_4$). First, a $l=20\,\mathrm{\mu m}$ long and $w=170\,\mathrm{nm}$ wide nanobeam as well as supporting clamping rectangles are defined using e-beam lithography and covered with aluminum serving as an etch mask. Subsequently, the unprotected silicon nitride and approximately 100\,nm of SiO$_2$  are removed by an anisotropic SF$_6$ reactive ion etching (RIE) step. In this way, we align the lower surface of the beam (i.\,e.~the SiO$_2${}-Si$_3$N$_4$ interface) approximately with the upper surface of the subsequently deposited niobium film (cf.~schematic displayed in Fig.~\ref{fig:grAssembly-CapacityVsBeamDisplacement}a). With a second e-beam lithography step followed by aluminum sputtering, we define a small rectangularly-shaped protective cover for the Si$_3$N$_4$ beam and its vicinity. Then, a $100\,\mathrm{nm}$ thick niobium film is deposited by magnetron sputtering. Subsequently, the microwave resonator and its input and output lines are patterned by a third e-beam lithography and a second RIE step. The resist and the aluminum coating are removed with acetone, potassium hydroxide and Piranha (i.\,e. a mixture of hydrogen peroxide and sulphuric acid). Finally, we release the silicon nitride beam using buffered hydrofluoric acid.

Using microwave transmission spectroscopy, we find a fundamental frequency of the superconducting microwave cavity of $\omega_c/2\pi=5.67\,\mathrm{GHz}$ and a linewidth of $\kappa/2\pi=749\,\mathrm{kHz}$, as shown in Fig.~\ref{fig:SEMimage-MicrowaveResonator}d.

\begin{figure}[tb]
\centering
\includegraphics[width=0.99\columnwidth]{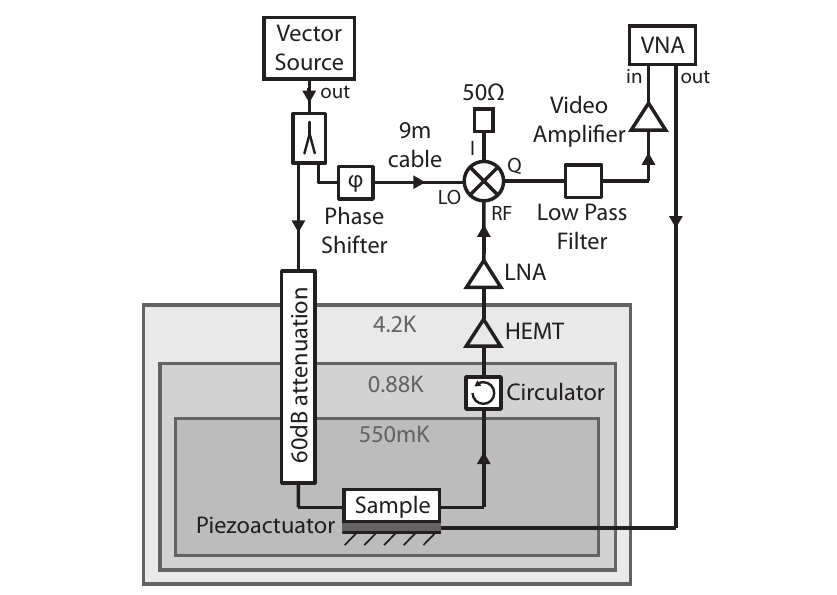}
\caption{Schematic of the experimental setup used to characterize the sample. The microwave signal from the vector source is split up into a drive tone for the superconducting microwave resonator and a reference tone. The drive tone is attenuated within the cryostat and capacitively coupled to the superconducting microwave resonator. The sample output signal is amplified, downconverted, filtered and detected with a VNA that drives the piezoelectric actuator on which the sample is mounted.}
\label{fig:CryoSetup-Homodyne-DWP-CalTone}
\end{figure}

To experimentally characterize the nano-electromechanical hybrid system, we employ the homodyne measurement setup depicted in Fig.~\ref{fig:CryoSetup-Homodyne-DWP-CalTone}. The microwave resonator is driven by a microwave source at its resonance frequency $\omega_c/2\pi$ with an estimated microwave power of $2.5\,\mathrm{nW}$ at the device input. The output microwave signal is amplified with a cold HEMT amplifier and a broadband low-noise amplifier (LNA). The sample is mounted on a piezoelectric actuator driven by the output of a vector network analyzer (VNA). This allows us to excite the fundamental flexural mode of the nanobeam to a high amplitude state. For phase sensitive detection with the VNA, the microwave signal is downconverted using an I-Q-demodulator. A phase shifter is used to adjust the phase such that the homodyne signal measured at the VNA is maximum corresponding to the quadrature response of the device.

When driving the microwave cavity and the beam simultaneously, the microwave drive tone with frequency $\omega_d/2\pi$ is modulated by the beam's motion, resulting in the generation of sidebands at $\omega_d\pm\Omega_{\mathrm{m}}$. The corresponding fluctuation amplitude is given by $\delta\omega_c=G\,x_0$ \cite{gorodetsky_determination_2010}, where $G$ is the electromechanical coupling and $x_0$ the mechanical amplitude. In the homodyne setup employed here, this frequency fluctuation $\delta\omega_c$ translates into the measured down-converted signal \cite{gorodetsky_determination_2010}
\begin{equation}\label{eq:PhomC}
P_{\mathrm{hom}} = \frac{K(\Omega)}{\Omega^2}\,\delta\omega_c^2 = \frac{K(\Omega)G^2}{\Omega^2} x_0^2
\end{equation}
with the oscillation frequency of the beam $\Omega$ and the transfer function $K(\Omega)$.

Figure~\ref{fig:SEMimage-MicrowaveResonator}e shows the down-converted spectroscopy signal of the silicon nitride nanobeam when excited in the linear (weak driving) regime. Here, the nanobeam follows the behavior expected for a harmonic oscillator. At $T\approx 550\,\mathrm{mK}$, we find an eigenfrequency $\Omega_{\mathrm{m}}/2\pi=13.95225\,\mathrm{MHz}$ of the nanobeam with a linewidth of $\Gamma_{\mathrm{m}}/2\pi=29\,\mathrm{Hz}$ corresponding to a quality factor of $Q=480{,}000$. This quality factor exceeds those of comparable nanobeams with Nb metallization \cite{hocke_phdthesis_2013} by more than a factor of three. From the geometry of the beam, we estimate the effective mass $m_{\mathrm{eff}}=0.43\,\mathrm{pg}$ (ref.~\citenum{seeSI}).

Upon increasing the drive power, nonlinear effects modify the dynamics of the beam \cite{timoshenko_vibration_1937}. In particular, the restoring force has to be modified by a cubic term leading to the Duffing equation of motion \cite{timoshenko_vibration_1937,nayfeh_nonlinear_1979}
\begin{equation}
\ddot{x}(t)+\Gamma_{\mathrm{m}}\dot{x}(t)+\Omega_{\mathrm{m}}^2 x(t)+\alpha x^3(t)=\frac{F}{m_{\mathrm{eff}}} \cos(\omega t)
\end{equation}
with the oscillating driving force $F\cos(\omega t)$. In case of a highly pre-stressed nanobeam, the Duffing nonlinearity $\alpha$ is given by \cite{unterreithmeier_coherent_2009}
\begin{equation}
\alpha=\pi^4\frac{E+\frac{3}{2}\sigma}{4l^2\rho} \; .
\end{equation}
Here, $E$, $\sigma$ and $\rho$ are the Young's modulus, the tensile stress and the mass density of the silicon nitride beam, respectively.

\begin{figure}[tb]
\centering
\includegraphics[width=0.99\columnwidth]{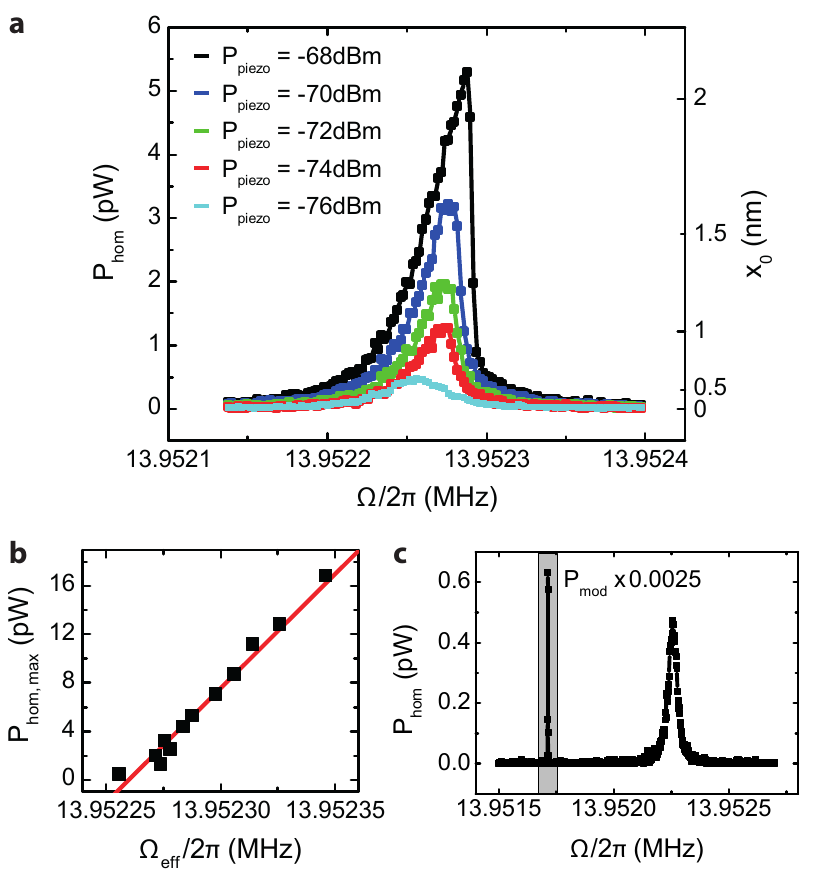}
\caption{\textbf{a.} Homodyne power spectrum (left axis) resp.~mechanical amplitude spectrum (right axis) of the nanobeam for varying external drive power. The amplitude scale (on the right) is based on the amplitude calibration described in the main text. \textbf{b.} Maximum homodyne power as a function of the effective resonance frequency (the so-called backbone curve). \textbf{c.} Homodyne power spectrum for low external drive ($P_{\text{piezo}}=-76\,\mathrm{dBm}$) using a frequency-modulated cavity drive tone. The sideband peak at $\Omega_{\text{mod}}/2\pi=13.9517\,\mathrm{MHz}$ is scaled by a factor of $0.0025$. Comparing the heights of both sideband peaks allows to determine the vacuum coupling $g_0$.}
\label{fig:DrivePowerSeries-BackboneCurve-FrequencyNoiseCalibration}
\end{figure}

The resulting amplitude spectrum of the Duffing oscillator is given by the implicit equation \cite{nayfeh_nonlinear_1979}
\begin{equation}\label{equ:duffing_amplitude_spectrum}
\left[\Gamma_{\mathrm{m}}^2+4\left(\Omega-\Omega_{\mathrm{m}}- \frac{3}{8}\frac{\alpha}{\Omega_{\mathrm{m}}}x_0^2\right)^2\right]x_0^2= \frac{F^2}{m_{\mathrm{eff}}^2\Omega_{\mathrm{m}}^2} \; .
\end{equation}
Thus, for increasing drive power the maximum of the amplitude spectrum, $x_{0,\mathrm{max}}$, shifts to higher frequencies $\Omega_{\mathrm{eff}}$, as illustrated in Fig.~\ref{fig:DrivePowerSeries-BackboneCurve-FrequencyNoiseCalibration}a. This dependence is described by the backbone curve \cite{nayfeh_nonlinear_1979}
\begin{equation}\label{AmplCal_backbone}
x_{0,\mathrm{max}}^2=\frac{8}{3}\frac{\Omega_{\mathrm{m}}}{\alpha}\left(\Omega_{\mathrm{eff}}-\Omega_{\mathrm{m}}\right).
\end{equation}
As a consequence, we can relate the amplitude $x_0$ of the beam to spectral information, which is straightforwardly accessible.

To determine the coupling between the microwave cavity and the nanobeam, we analyze the quadrature response of the superconducting microwave cavity. To this end, we calibrate this response with a known frequency modulation $\Omega_{\mathrm{mod}}$ of the microwave carrier frequency $\omega_d$ (refs.~\citenum{zhou_slowing_2013,gorodetsky_determination_2010}) as shown in Fig.~\ref{fig:DrivePowerSeries-BackboneCurve-FrequencyNoiseCalibration}c (highlighted in gray). Hereby, we can relate the detected power of the homodyne signal $P_{\mathrm{hom}}$ to the well-known frequency modulation depth, or in other words, we can determine the transfer function $K(\Omega_{\mathrm{mod}})$. We choose $\Omega_{\mathrm{mod}}\approx \Omega_{\mathrm{m}}$ and therefore assume $K(\Omega_{\mathrm{mod}})\approx K(\Omega_{\mathrm{m}})$ \cite{zhou_slowing_2013,gorodetsky_determination_2010}.

Using Eq.~(\ref{eq:PhomC}), the backbone curve Eq.~(\ref{AmplCal_backbone}) reads
\begin{equation}\label{AmplCal_backbone2}
P_{\mathrm{hom,max}}=\frac{8}{3}\frac{\Omega_{\mathrm{m}}}{\alpha}\frac{K(\Omega_{\mathrm{m}})G^2}{\Omega_{\mathrm{m}}^2}\left(\Omega_{\mathrm{eff}}-\Omega_{\mathrm{m}}\right)
\end{equation}
where $P_{\mathrm{hom,max}} \equiv P_{\mathrm{hom}}(\Omega_{\mathrm{eff}})$ denotes the down-converted signal power at the effective resonance frequency $\Omega_{\mathrm{eff}}$, that is, at the maximum of the $P_{\mathrm{hom}}(\Omega)$ curve. Thus, measuring the homodyne power spectrum of the beam as a function of the external driving force imposed by the piezoactuator and plotting $P_{\mathrm{hom,max}}$ versus $\Omega_{\mathrm{eff}}$, as shown in Fig.~\ref{fig:DrivePowerSeries-BackboneCurve-FrequencyNoiseCalibration}b, allows to extract the coupling constant $G$. With $K(\Omega_{\mathrm{m}})=2.4\times 10^{-3}\,\mathrm{W}$, $E=160\,\mathrm{GPa}$ (ref.~\citenum{unterreithmeier_damping_2010}), $\sigma=830\,\mathrm{MPa}$ (ref.~\citenum{unterreithmeier_damping_2010}), $\rho=2600\,\mathrm{kg/m^3}$ (ref.~\citenum{faust_microwave_2012}) and $m_{\mathrm{eff}}=0.43\,\mathrm{pg}$, we obtain $G/2\pi=312\,\mathrm{Hz/nm}$. Thus, the electromechanical vacuum coupling is $g_0/2\pi=G/2\pi\cdot x_{\mathrm{zpf}}=11.5\,\mathrm{mHz}$ (ref.~\citenum{gorodetsky_determination_2010}) with the zero-point fluctuation of the beam $x_{\mathrm{zpf}}=\sqrt{\hbar/2m_{\mathrm{eff}}\Omega_{\mathrm{m}}}=37\,\mathrm{fm}$ (ref.~\citenum{kippenberg_cavity_2008}). Compared to similar nano-electromechanical hybrid systems with metallized beams (see e.g.~ref.~\citenum{zhou_slowing_2013}), the coupling is about two orders of magnitude smaller, as it solely relies on the dielectric interaction between nanobeam and niobium electrodes.

To understand the electro-mechanical coupling mechanism in more detail and to estimate the coupling constant a priori, we use a COMSOL 2D model representing the cross-sectional sample area perpendicular to the longitudinal nanobeam direction (see Fig.~\ref{fig:grAssembly-CapacityVsBeamDisplacement}a, cf.~ref.~\citenum{unterreithmeier_universal_2009}). Hereby, we calculate the capacitance per length, $c_g$, between center line and ground plane for various (static) displacements $x_0$ of the nanobeam. The result is plotted in Fig.~\ref{fig:grAssembly-CapacityVsBeamDisplacement}b. We find a capacitance change per displacement of $\partial c_g/\partial x_0=2.4\,\mathrm{(fF/m)/nm}$. Due to the vibrating nanobeam the total capacitance $C$ of the microwave resonator has two contributions: a static part $C_0$ and a displacement-dependent oscillating part $C_g(x_0)=c_g(x_0) l_{\mathrm{eff}}$, where $l_{\mathrm{eff}} =0.613 l$ is the effective length of the beam \cite{seeSI}.

The resonance frequency of a superconducting $\lambda/2$ coplanar waveguide resonator is given by $\omega_c=1/\sqrt{LC}$, where $L$ denotes the inductance of the resonator \cite{goeppl_coplanar_2008}. Thus, assuming $C_g/C_0 \ll 1$, the resonance frequency can be rewritten as
\begin{equation}
\omega_c(x_0)\approx \omega_0\left(1-\frac{C_g(x_0)}{2C_0}\right),
\end{equation}
where we have introduced the resonance frequency of the microwave cavity for an undisplaced beam, $\omega_0=1/\sqrt{LC_0}$. With the line impedance $Z_0$, we obtain the coupling constant
\begin{equation}
G=-\frac{\partial \omega_c}{\partial x_0} = \frac{Z_0\omega_0^2}{\pi} \frac{\partial C_g}{\partial x_0}.
\end{equation}
Using the numerically obtained value $\partial c_g/\partial x_0=2.4\,\mathrm{(fF/m)/nm}$, the measured resonance frequency of the microwave cavity $\omega_0/2\pi=5.67\,\mathrm{GHz}$ and its designed impedance value $Z_0=70\,\mathrm{\Omega}$, we obtain $G/2\pi=132\,\mathrm{Hz/nm}$ or $g_0/2\pi=4.8\,\mathrm{mHz}$. Although the simulation uses only an approximation of the real sample geometry, this is in good agreement with the experimentally determined value, demonstrating that the applied modeling is useful for an a-priori estimation of the nano-electromechanical coupling of a given sample geometry.

In conclusion, we have fabricated and characterized a nano-electromechanical hybrid system consisting of a superconducting microwave resonator and a non-metallized nanomechanical beam. On decreasing the temperature, for a beam with a resonance frequency of about $14\,\mathrm{MHz}$ we observe an increase of the mechanical quality factor from $40{,}000$ at room temperature \footnote{The room temperature quality factor of the nanobeam has been determined by optical interferometry.} to $480{,}000$ at $550\,\mathrm{mK}$. For strong external driving forces we observe the transition from the linear to the Duffing regime and use this to quantify the mechanical displacement amplitude of the beam. Fitting the peak values of the measured homodyne power spectra allows us to precisely determine the mechanical amplitude. This method is complementary to the usually employed calibration via thermal motion \cite{hauer_general_2013,gorodetsky_determination_2010} and especially useful for systems where the Brownian motion is not straightforwardly detectable. Moreover, the precise knowledge of the motional amplitude allows us to derive the electromechanical coupling $g_0$. For our device, we find $g_0/2\pi=11.5\,\mathrm{mHz}$, which is corroborated by numerical modeling of the device.

This work opens the path for further experimental studies of mechanical losses in silicon nitride at millikelvin temperatures, extending previous work on the damping mechanisms in Si$_3$N$_4${} nanomechanical beams \cite{unterreithmeier_damping_2010,faust_signatures_2014}. Moreover, the concept of dielectrically coupling a pure Si$_3$N$_4${} nanobeam to a high-$Q$ microwave resonator is promising especially for sensing devices, e.\,g.~for the detection of single molecules, which require high frequency resolution and thus low damping constants.

Financial support by the Deutsche Forschungsgemeinschaft via Project No. Ko 416/18 is gratefully acknowledged.


%

\pagebreak
\widetext
\vspace{1cm}
\begin{center}
\textbf{\large Supplemental Material: Circuit Electromechanics with a Non-Metallized Nanobeam}
\end{center}

\renewcommand{\thetable}{\Roman{table}}
\renewcommand{\thesection}{\Roman{section}}
\renewcommand*{\citenumfont}[1]{S#1}
\renewcommand*{\bibnumfmt}[1]{[S#1]}

\makeatletter
\makeatother

\renewcommand{\thesection}{\Alph{section}}
\renewcommand{\thesubsection}{\alph{subsection}}
\renewcommand{\thefigure}{S\arabic{figure}}
\renewcommand{\thetable}{S\arabic{table}}
\renewcommand{\theequation}{S\arabic{equation}}
\renewcommand{\refname}{Additional References}

\setcounter{equation}{0}


\section{Effective mass of the beam}
According to the Supplementary Information of ref.~\citenum{unterreithmeier_damping_2010}, the total energy of the oscillating beam is given by
\begin{equation}\label{SI_U}
U=\frac{1}{2}\rho w t \Omega_{\mathrm{m}}^2\int_{-l/2}^{l/2} dz\,x_0^2(z),
\end{equation}
where $x_0(z)$ is the $z$-dependent displacement of the beam, $\rho$ and $\Omega_{\mathrm{m}}$ are the density of the beam resp.~its resonance frequency, and $w$, $t$ and $l$ denote width, thickness and length of the beam.

Following Euler-Bernoulli beam theory \cite{timoshenko_vibration_1937,hocke_exploring_2014_S}, the displacement $x_{0}(z)$ of the fundamental mode of the presented tensile-stressed doubly-clamped beam is given by
\begin{equation}\label{eq:SI_x0n}
x_0(z)=a_1 \mathrm{e}^{\alpha z} + a_2 \mathrm{e}^{-\alpha z} + a_3 \sin(\beta z) + a_4 \cos(\beta z)
\end{equation}
with the numerically calculated coefficients $\alpha = 2.50\times 10^6\,\mathrm{m^{-1}}$, $\beta = 1.63\times 10^5\,\mathrm{m^{-1}}$, $a_1 = a_2 = 9.04\times 10^{-13} a_4$ and $a_3=-3.30\times 10^{-13} a_4$.

With that, the integral in (\ref{SI_U}) can be solved numerically:
\begin{equation*}
\int_{-l/2}^{l/2} dz\,x_0^2(z)=0.480\,l\,a_4^2
\end{equation*}

We define the effective mass $m_{\mathrm{eff}}$ of the beam by
\begin{equation}\label{SI_U2}
U=\frac{1}{2}m_{\mathrm{eff}}\Omega_{\mathrm{m}}^2 x_{0,\mathrm{c}}^2
\end{equation}
where $x_{0,\mathrm{c}}:=x_0(z=0)$ is the displacement of the center of the beam.

Comparing Eqs.~(\ref{SI_U}) and (\ref{SI_U2}), we get
\begin{equation*}
m_{\mathrm{eff}}=\frac{\rho w t}{x_{0,\mathrm{c}}^2}\int_{-l/2}^{l/2} dz\,x_0^2(z)=0.480\,\rho w t l=0.43\,\mathrm{pg}
\end{equation*}

\section{Effective length of the nanobeam}
As described in the main text, a static 2D COMSOL model is employed to calculate the effect of the beam displacement on the capacitance between center line and ground plane of the microwave resonator. Considering that the displacement $x_0$ is a function of the position $z$ along the beam, the capacitance variation $\delta C_g$ induced by a displacement $x_0(z)$ is given by
\begin{equation*}
\delta C_g=\frac{\partial c_g}{\partial x} \int_{-l/2}^{l/2} dz\,x_0(z)
\end{equation*}
In linear approximation, we write
\begin{equation*}
\frac{\partial C_g}{\partial x}\approx \frac{\delta C_g}{x_{0,\mathrm{c}}}=\frac{\partial c_g}{\partial x} \int_{-l/2}^{l/2} dz\,\frac{x_0(z)}{x_{0,\mathrm{c}}}
\end{equation*}

As above, the integral can be solved numerically using (\ref{eq:SI_x0n}):
\begin{equation*}
\int_{-l/2}^{l/2} dz\,\frac{x_0(z)}{x_{0,\mathrm{c}}}=0.613\,l
\end{equation*}

Thus, we have
\begin{equation*}
\frac{\partial C_g}{\partial x}=\frac{\partial c_g}{\partial x} \cdot l_{\mathrm{eff}}
\end{equation*}
where $l_{\mathrm{eff}}=0.613\,l$ is the effective length of the beam.

\end{document}